# Quantitative and bond-traceable resonant X-ray optical tensors of organic molecules

Victor Murcia[1], Obaid Alqahtani[1,2], Harlan Heilman[3], Brian A. Collins[1,3*]

[1]*Department of Materials Science and Engineering, Washington State University, Pullman, Washington 99164, USA*

[2]*Department of Physics, Prince Satam bin Abdulaziz University, Alkharj 11942, KSA*

[3]*Department of Physics and Astronomy, Washington State University, Pullman, Washington 99164, USA*

Local molecular orientation and conformation within nanostructures is central to their behavior in devices, emergent quantum states, and biological processes. X-ray microscopy and scattering at the carbon absorption edge is uniquely sensitive to these phenomena, through transition dipole moments encoded with photon energy and polarization. However, quantitative analysis is precluded due to the lack of accurate optical models with bond and orientation specificity. We generate such a model through an unsupervised machine learning algorithm that clusters density functional theory calculations into principle electronic transitions, quantitatively refined with angle-resolved absorbance measurements. The resulting optical tensor reproduces data from samples with domains of different orientation and crystalline packing. This refinement method connects spectral features to chemical moieties and resolves measurement-inaccessible dipole moments, establishing a new paradigm for label-free orientation analyses of bonds within molecular nanostructures.

Molecular orientation, alignment and conformation is central to the functional properties and behaviors of molecular materials especially in organic devices and biological materials. For example, molecular orientation governs emission efficiency in organic light emitting diodes [1,2], charge transport in printable transistors [3,4], charge generation in solar cells [5], doping dynamics in ionic blends [6], self-assembly in supramolecular nanostructures [7,8], and quantum states and dynamics. [9–11]

X-rays resonant with electronic transitions between molecular orbitals are sensitive to the orientation of a molecule's chemical moieties through photon polarization. Linear dichroism in near edge X-ray absorption fine structure (NEXAFS) spectroscopy can be used to characterize a sample's global molecular orientation. [12] More recently sensitivity to *local* molecular orientation has been demonstrated through resonant scattering, revealing orientation within nanostructures and 3D interfaces, [13,14] for example, chiral ordering in liquid crystals, [15–17] and thin film interfacial substructure. [18,19]

While these studies have revealed orientational ordering that is inaccessible by any other means, they remain either highly qualitative or involve simplistic dielectric models with unknown validity. In particular, the best models are expanded from angle-resolved NEXAFS spectroscopy measurements where the molecular origin of spectral features is assigned via comparison to that of similar compounds. [12,20–22] However, there typically exists numerous spectral features at the carbon edge due to multiple carbon atoms in a macromolecule such that the chemical origin cannot be validated. Furthermore, absorption measurements average out orientation due to low crystallinity such that they cannot determine the transition dipole moment (TDM) orientation of spectral features. Variations in molecular conformation further confuse the results. [19,23] This limits the accuracy of bond identification and orientation analyses and prevents expanded analysis to more complex situations such as molecular conformation in biological assemblies.

Density functional theory (DFT) calculations have been employed to identify the chemical origins of spectral features in spectroscopy studies [24–26] and more recently, studies involving resonant scattering. [27–29] Unfortunately, these calculations lack accuracy with only qualitative resemblances due to limitations of the DFT basis sets, functionals, and methods [30,31], leading to incorrect predictions for transition energies and oscillator strengths. These inaccuracies result in subjective interpretation and assignment of spectral features, limiting the accuracy of absorption analyses. This problem is compounded in scattering where oscillator strength errors simultaneously affect both the absorption and phase of the optical response. Furthermore, with so many carbon atoms in each molecule, DFT calculates too many transitions to be refined to experimental data.

Machine learning (ML) methods have emerged as powerful tools for interpreting X-ray absorption spectra, enabling automated feature extraction, prediction of spectral shapes, and improved assignment of local bonding motifs [32–35]. Recent work demonstrates that ML can overcome some limitations of first-principles methods, but these approaches still do not provide bond-resolved optical models needed for quantitative orientational analysis. Unsupervised clustering algorithms [36–38], are widely used to identify structure within datasets and reduce dimensionality. They offer an opportunity to bridge the gap between experiment and theory by organizing the dense manifold of DFT-generated transitions and enabling a traceable refinement of DFT calculations to experimental data.

Here we demonstrate an interpretable clustering algorithm that reduces the dimensionality of transition potential (TP-)DFT-simulated NEXAFS spectra [31,39] of a molecule that retains the principle features. We use this to create an optical model that encodes the missing information from a NEXAFS measurement (i.e. chemical origin and TDM orientation) while the reduced dimensionality enables stable experimental refinement of DFT oscillator strengths and transition energies. The refined optical model quantitatively matches experimental NEXAFS spectra as well as identifies the chemical origins of experimental spectral features originating from unique moieties within the molecule. The model is validated against experimental data of the same molecules but with different orientation and crystalline packing.

Copper (II) phthalocyanine (CuPc) was used for this study due to its organometallic nature providing bond variety to challenge DFT calculations, while exhibiting high crystallinity, a rigid conformation, and $D_{4h}$ symmetry, allowing for a uniaxial optical model. CuPc is, additionally, a foundational molecular system across multiple scientific communities. It is widely characterized and implemented into a variety of organic electronic applications like photovoltaics [40,41], light emitting diodes [42] and transistors [43]. The electronic configuration and assembly of Pc molecules have also been a focus in quantum phenomena such as Andreev reflection and the Kondo effect. [9–11] Pc molecules, furthermore, have a similar molecular framework and electronic orbitals to chlorophyl and hemoglobin – central molecules to plants and animals, respectively.

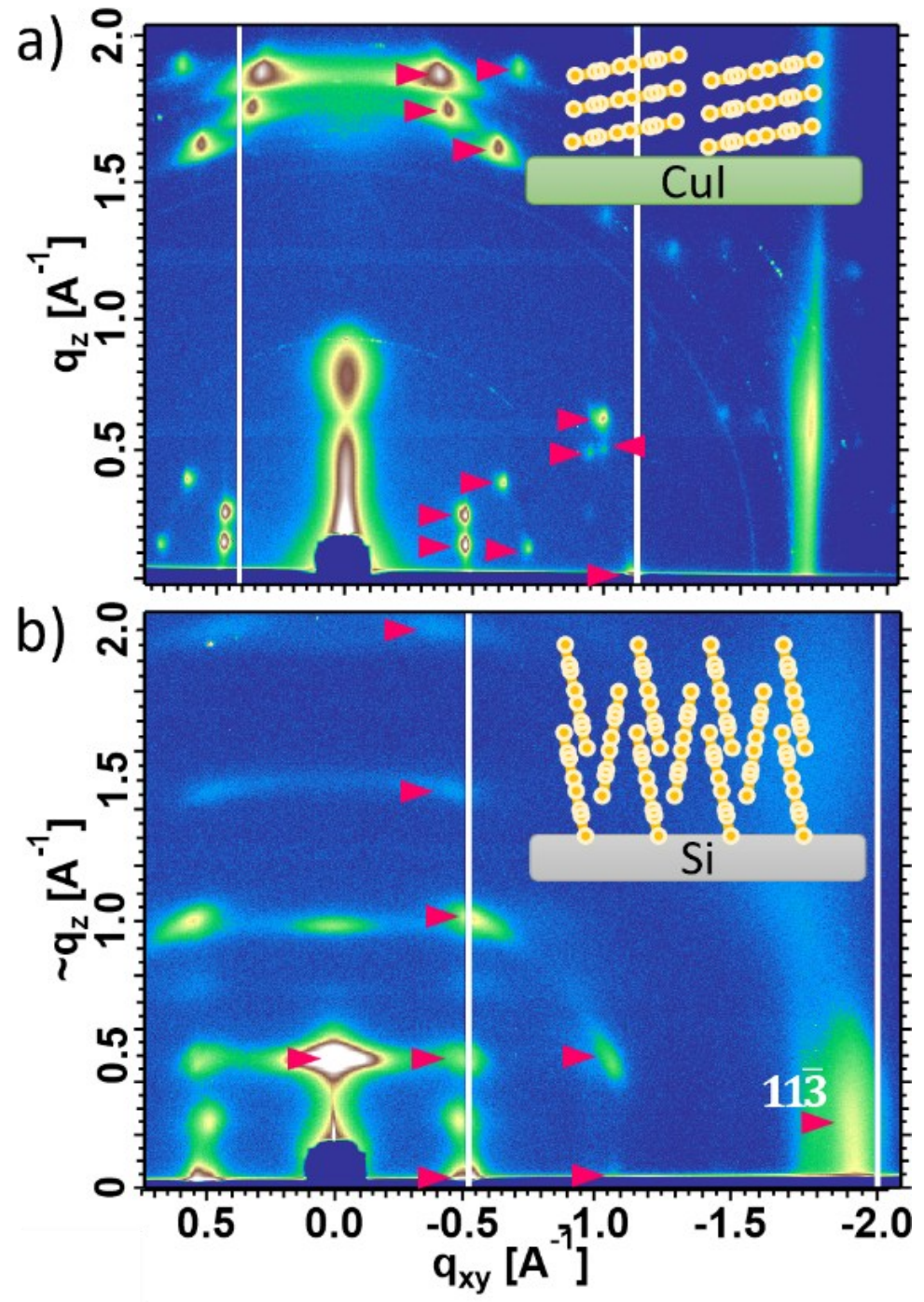

*Fig. 1. GIXRD demonstrating packing and orientation modulation of CuPc films for a) CuPc on CuI and b) CuPc on bare Si. Red markers point to Bragg reflections identified in references cited in the text. Inset are side-view schematics of the packing structure identified.*

Modulating CuPc orientation and packing into herringbone and brickstone [44–46] structures is readily accomplished via substrate choice [47,48] and verified in **Fig. 1** via grazing incidence X-ray diffraction (GIXRD) measurements. In particular, the uniquely defined diffraction patterns identify successful growth of the polycrystalline brickstone $(11\bar{2})$ orientation (**Figure 1a**) [45,47] and herringbone (200) orientation (**Figure 1b**) [46,47]. This enables one sample to be used to refine the optical model while the other sample tests the model's robustness to varying orientation and crystal structure.

**Fig. 2a and b** show the atomic identities and positions from detailed XRD measurements [49] and the DFT-simulated NEXAFS for CuPc compared to experiment with good qualitative agreement across the entire near edge energy range. However, inaccuracies in transition energy and oscillator strengths preclude a quantitative match, and thousands of transitions (sticks in **Figure 2b**) prevents refinement to the experimental data. **Figure 2c** shows a zoom of the pre-edge energy range with transitions color coded by chemically unique excitation center. Notably, transitions can be qualitatively clustered as we have subjectively labeled into spectral features. Interestingly, DFT features A and C lack contributions from the imine carbons and also exhibit systematically low oscillator strengths relative to experimental measurements. DFT features D and E exhibit inaccurate energies with E mainly originating from only the imine carbons. These inaccuracies could originate from inaccurate functionals or the nature of overlap with the final states (unoccupied molecular orbitals, MOs, described below).

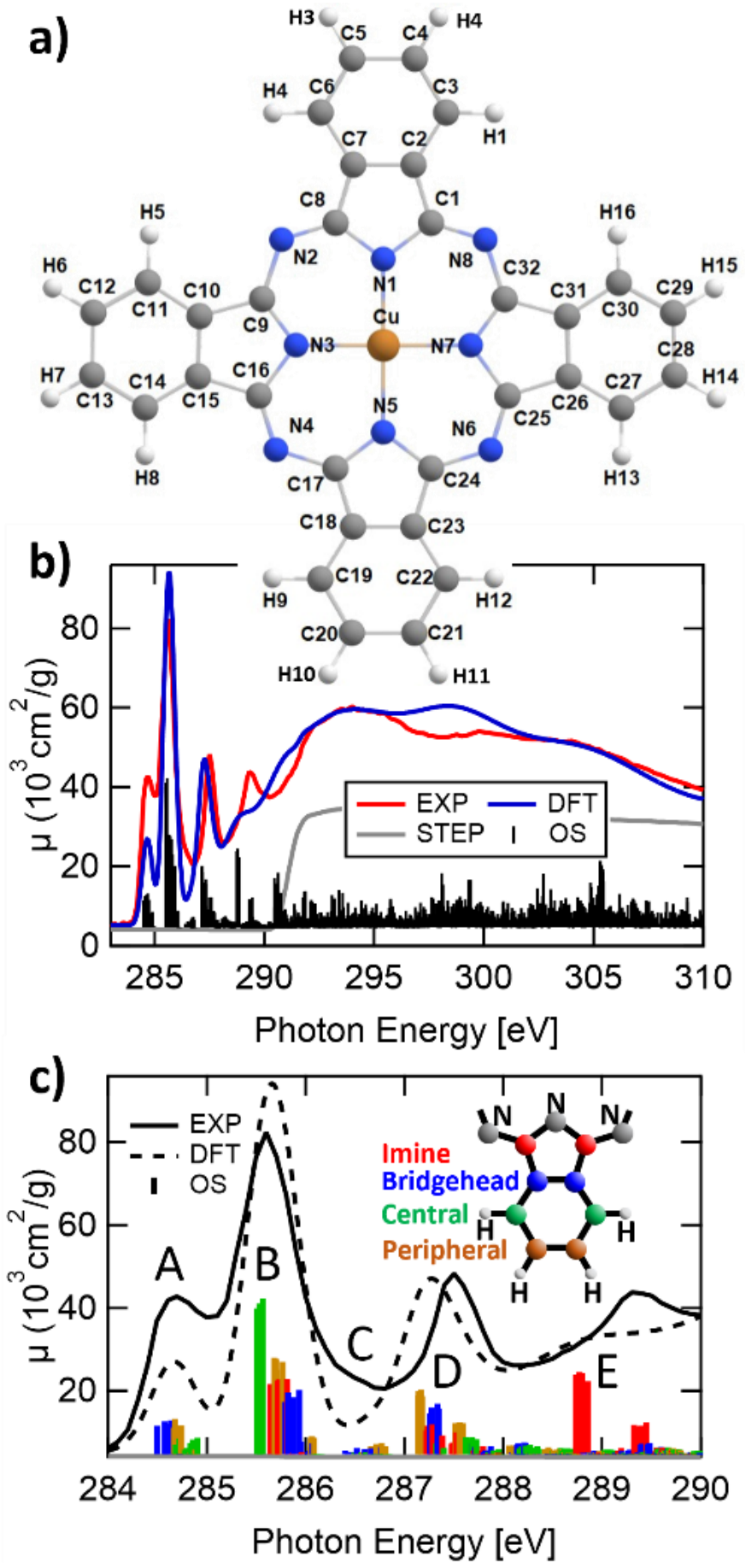

*Fig. 2. TP-DFT calculations compared with experiment. a) Molecular structure of CuPc used to calculate the NEXAFS. b) NEXAFS spectra derived from experiment (red, at the magic angle) and DFT calculations (blue) that includes the calculated photoionization function (grey). Black vertical lines encode transition oscillator strength. c) Zoom of (b) to the pre-edge region with some clusters labeled. Transitions are color-coded by chemically unique carbon atom labeled in the inset schematic.*

To make quantitative refinement of DFT calculations manageable, we developed an unsupervised ML algorithm that reduces the dimensionality of the DFT parameter space while simultaneously preserving the integrity of the calculated features. The algorithm entails two steps: (1) filtering of transitions with an insignificant contribution to the spectrum below an oscillator strength threshold (OST) and (2) an agglomerative hierarchical clustering of spectrally similar transitions above an overlap threshold (OVPT). For the clustering step, each transition is represented by a gaussian peak defined by transition energies, oscillator strengths and a linear, energy-dependent broadening scheme. The percent overlap area of these peaks serves as our distance metric and is computed for each transition pair, generating an overlap matrix as shown in **Fig. 3a**. Using peak overlap area (example inset) as the distance metric is appropriate for our system because the clustering occurs entirely in the energy space, making it a physically meaningful measure of spectral proximity. However, this methodology is extensible to other distance metrics between transitions. Overlap areas greater than the OVPT are clustered and represented as a new single peak.

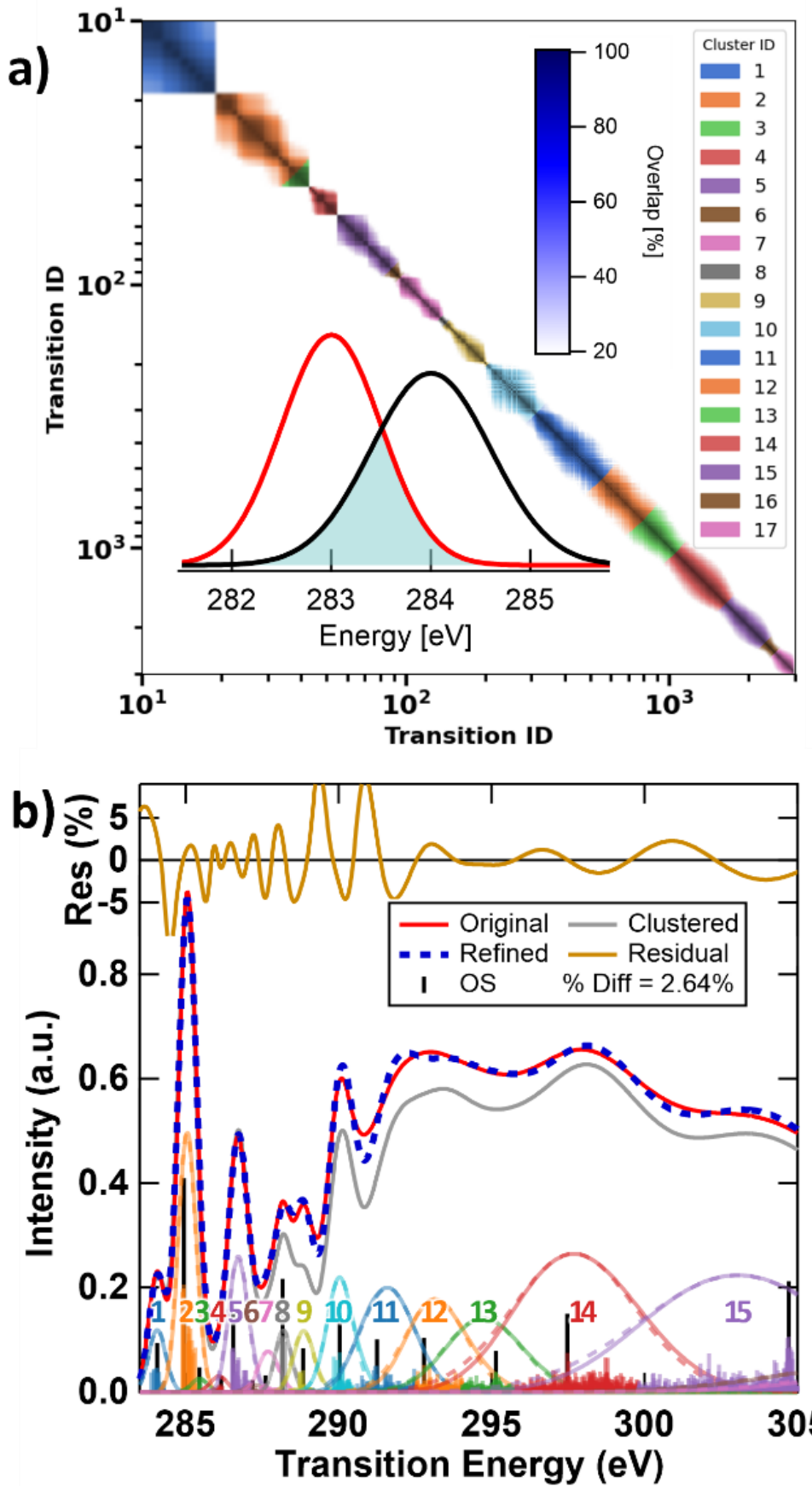

*Fig. 3. Results of clustering algorithm on TP-DFT calculation. a) Overlap matrix for the OS filtered transitions. Final clusters are color-coded with darkness representative of the peak overlap. Inset: The overlap area between two example transistions. b) The clustered NEXAFS transitions (grey) are amplitude-fit (blue dash) to the original DFT (red) with residuals (gold). Each cluster (peak) and associated transitions (sticks) are color-coded to the overlap matrix in (a). Solid lines represent the final gaussian while dashed lines respresent the distribution of intensity from the sum of individual transitions.*

The overlap values are shown in **Fig. 3a**, color-coded by their final clusters. Here, transitions to a new cluster are revealed by a characteristic narrowing of off-diagonal elements that contain significant overlap. **Fig. 3b** shows the result of the full algorithm compared with the original calculated spectrum. DFT calculated transitions and their ML clustered state is provided in **SI file 1** [50]. The OST filtering lowers the overall intensity (gray trace), so the clusters are amplitude-fit to the original DFT (dashed blue line). An excellent agreement with the original calculation results (average deviation 2.6% with worst residuals <10%, gold line), while being described by 763x fewer parameters. The final cluster peaks (**Fig. 3b bottom**) encompass their transitions (stick markers) and are generally well-described by an OS-dominant principal transition (black sticks) near the center of each cluster.

The clustered DFT model can now be refined to experimental data, resulting in a quantitatively accurate optical model based on DFT. We accomplish this by fitting the cluster peak amplitudes, positions, widths, and TDM orientation to angle-resolved NEXAFS spectroscopy measurements of the brickstone-structured CuPc film. One additional fit parameter was the frame of reference rotation from the molecular frame to the sample frame known as the averge molecular tilt (polar) angle $\alpha$. The level of success is determined by the reduced chi-squared $\chi^2$ and the extent to which the parameters must be refined. Since the two input parameters (OST and OVPT) will change the outcome of the algorithm, we systematically studied a range of their values, plotting the final $\chi^2$ in **Figure 4a**. The fit becomes poor below an OPVT of 40% and above 70%, where the lower threshold combines transitions into too few clusters which can't reproduce the experimental spectral features. Higher OPVTs result in too many clusters (parameters), making the fit unstable. Generally higher OST values also results in poorer fits. Multiple sets of parameters result in a similarly high quality fit to the experimental data ($\chi^2 < 1.3$).

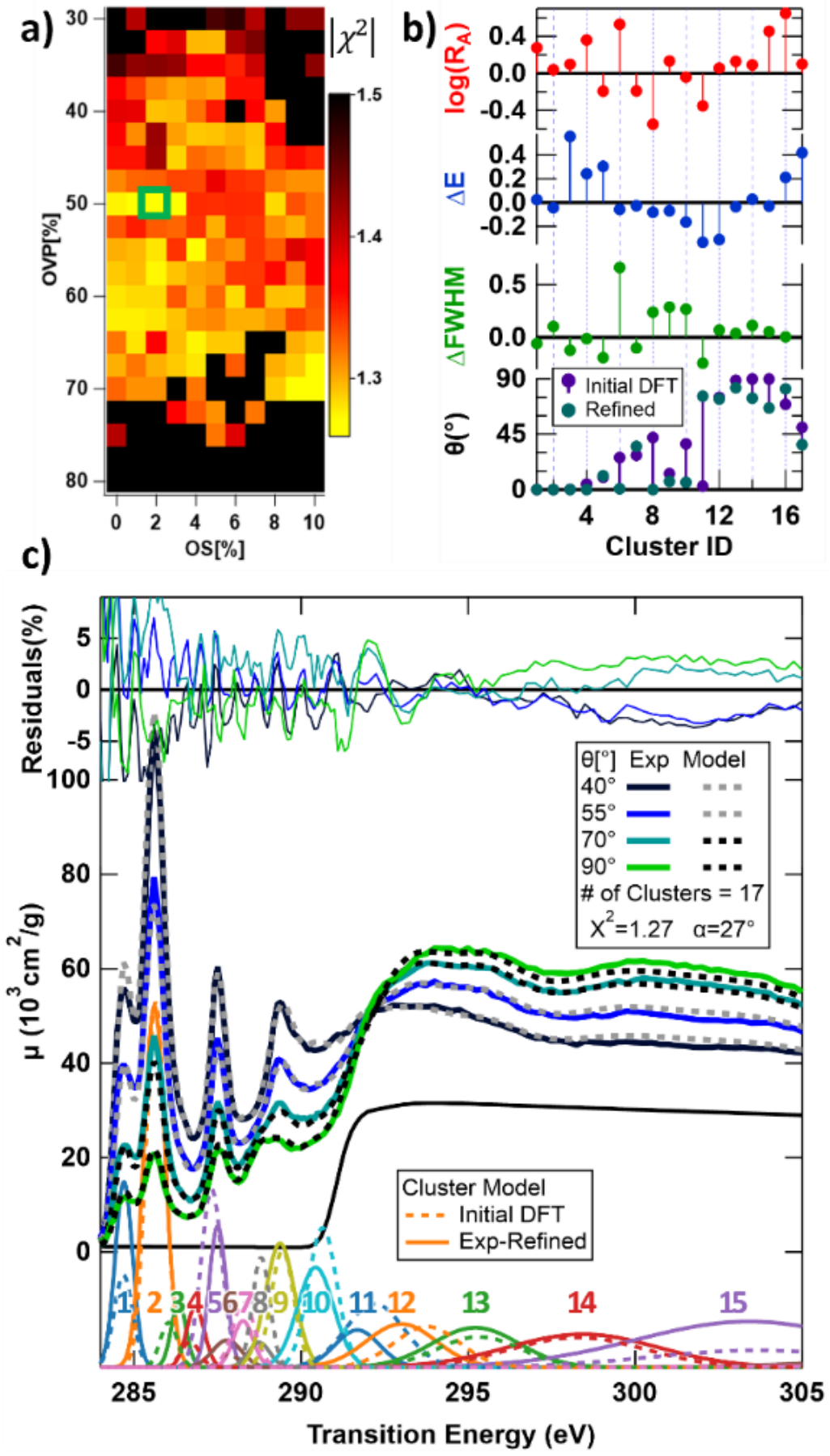

*Fig. 4. Determining best CuPc optical model a) $\chi^2$ versus OST & OPVT parameters. The chosen optimal parameter set is framed in green at 2% OST and 50% OVPT. b) Change in the parameters from the refinment fit of the DFT-clusered model to the experimental angle-resolved NEXAFS of the brickstone CuPc film. [51] Top Down: Log ratio of the cluster amplitudes (fit/initial: $R_A = A_f/A_i$); Position change normalized to original cluster FWHM; width change; and change in TDM polar angle. c) Fit results, residals, and individual cluster changes (offset for clarity) for the final refined optical model.*

The optimal parameter set was deteremined to be OST=2% and OVPT=50% (green square in **Figure 4a**). This was based not only on low $\chi^2 = 1.27$ but also minimizing how much the parameters needed to be refined to fit experimental data shown in **Figure 4b** with other candidate sets compared in **Figure S7**.

Overall, refinements are small with all cluster positions staying within the FWHM of their original position and almost no changes in cluster widths. The largest refinements are to amplitudes (oscillator strengths) and to a lesser extent to TDM orientations. The final refined model is shown in **Figure 4c** which also displays the cluster gaussians before and after refinement to the experimental data. The residuals show that the refined cluster model follows the experimental data to within 5% at energies above the first two clusters. This plus slighly higher residuals in the first two clusters likely originates from the error of the measurement in following the theoretical $I \propto \cos^2\theta$ dependence (see **Figure S8**). Further supporting the accuracy of the refinement process is that the DFT frame of reference rotation $\alpha = 27°$, in agreement with that of the brickstone structure oriented on H-diamond and graphene [52,53]. The parameters of the final clustered optical tensor are provided in **SI File 2** [50].

The cluster peak functions plotted in **Figure 4c** show how little the DFT clusters were refined to quantitatively reproduce the experimental measurements. The most changed clusters (**Figure 4b**) either represent minor features (clusters 6 and 8), get moved across the isosbetic point (cluster 11), or are high energy peaks (clusters 15 & 16). Small inacuracies in DFT-calculated orbital overlap can affect minor transitions, while errors in the ionization energy combined with a lack of exact exchange in the chosen functional may shift transitions across the edge. Finally, the amplitude changes of high energy clusters likely originate from well documented innacuracies of TP-DFT at describing many body effects and the continum state. [54–57] We also note that the present DFT framework does not explicitly capture vibrational effects, structural disorder, or multielectron excitations, which can contribute to deviations in the observed line shapes. [25,58] Despite these inacuracies, most of the clustered DFT model is not significantly changed by refinement to the experiment, preserving the physical attributes from the original calculation.

Another powerful aspect of the algorithm is that the final state MOs can be tracked and used to identify the chemical nature of the features in the measured spectra. **Figure 5a** shows the primary MOs within several of the clusters. The first three clusters reveal the C=C $\pi$-character of the lowest energy couplet feature in the spectra with the electron density strattling the atomic plane. Of note, the electron density of the dominant peak (cluster 2) avoids the metal center, likely making this feature invariant to different elements that occupy this location. Clusters 4 and 5, which encompass the feature at 287.5 eV, differentiate themselves by having electron density segregated to the benzene petals of the molecule, even avoiding the imine groups in the core to some extent. Cluster 7 is almost the inverse where the electron density focuses primarily on the metal-center. The TDM orientation of this cluster is calculated to be $\theta_c = 27.9°$ experimentally refined to $\theta_r = 35.1°$, which can be categorized in neither of the traditional categories of vector-like ($\theta \approx 0°$) nor planar-like ($\theta \approx 90°$) orbitals [12] – potentially due to the d-orbitals of the heteroatom metal center. By contrast, cluster 11 uniquely contains electron density segregated primarily on the imine groups only. The clusters at higher energies (for example cluster 13) have sigma character, with electon density in the region between atomic nuclei. Thus, our DFT-refinement algorithm, is able to connect specifc experimental spectral features to unique chemical moeities within the molecule – a capability not possible previously.

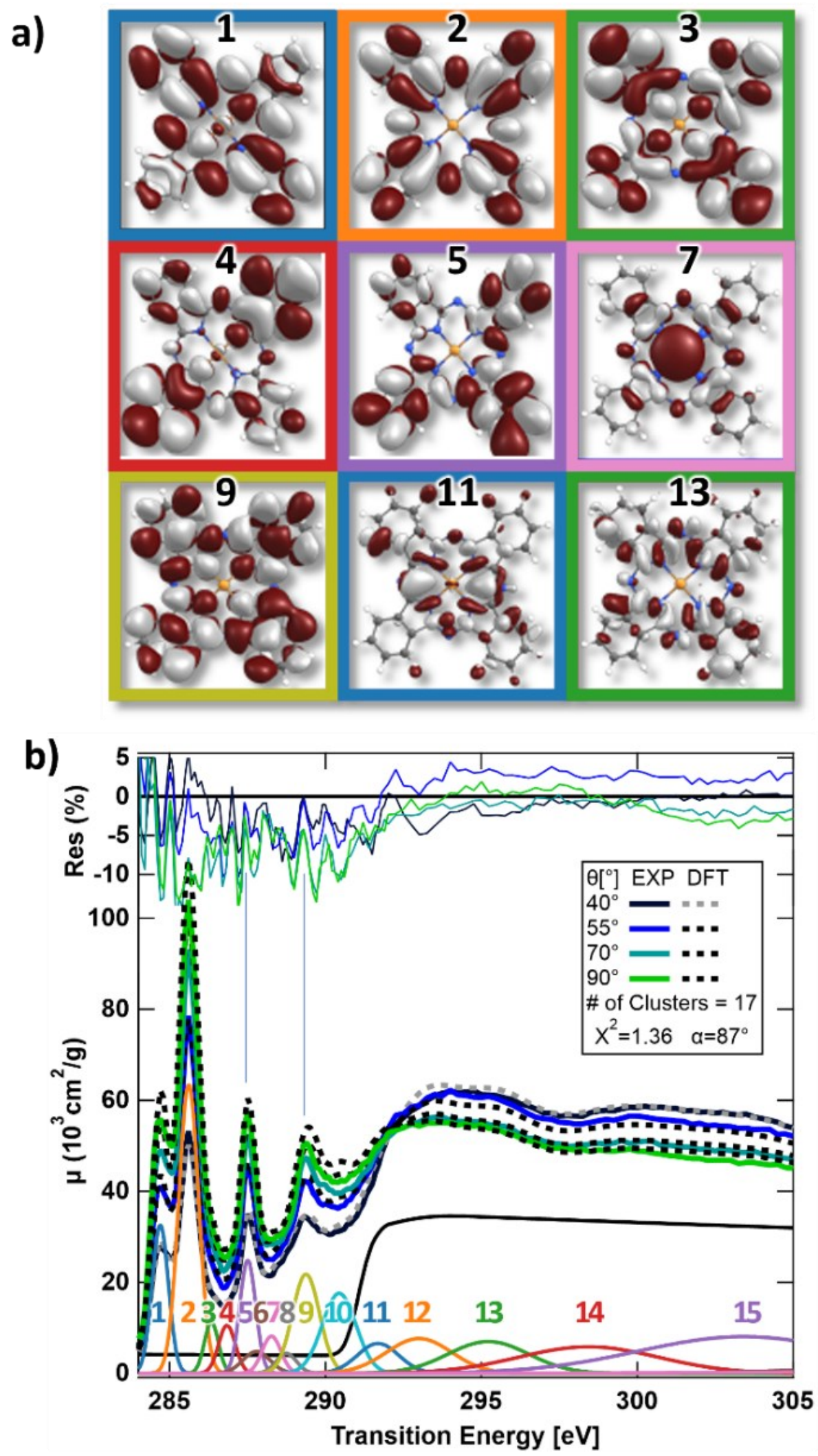

*Fig. 5. Chemical and orientational traceability of the new optical tensor. a) Molecular orbittals selected due to having the strongest OS within a cluster. b) The same model can quantitatively reproduce angle-resolved NEXAFS measurements on herringbone CuPc [59] via a single fit parameter representing average molecular orientation.*

We finally validate this model through its ability to reproduce data from the molecule in different physical states via our second CuPc sample crystalized into a different packing structure and different average orientation. **Figure 5b** displays the result of this analysis where we fit our refined optical model to angle-resolved NEXAFS spectra with only a single fit parameter: the average molecular tilt angle $\alpha$. The model is an excellent fit with $\chi^2 = 1.36$ with residuals again originating primarily from measurement artifacts shown in **Figure S9.** Importantly, the fitted tilt angle from this analysis $\alpha = 87.0 \pm 3.4°$ is consistent with the literature values [49,60] further supporting the validity of the final refined optical model.

This method for generating molecular optical models opens up several opportunities. First, it enables quantitative analyses of bond identity and orientation within a nanostructure from polarized scattering experiments where current models are often only qualitative or based on inferences. [61] Notably NEXAFS spectroscopy measurements are incapable of measuring local structures and are underdetermined due to population averaging, requiring simplistic 'vector/planar-like' TDM models. [12] The DFT-based optical model developed here resolves low-symmetry TDM orientations, for example, cluster 7 oriented at $\theta = 35°$ from the molecular polar axis that can be leveraged to solve nanostructures with unique orientations. Perhaps even more powerful is the ability to localize the chemical moieties involved in the transition at a specific X-ray energy. Similar to Raman spectroscopy or $^1$H NMR, this chemical specificity coupled with orientation sensitivity, could be used to identify specific macromolecular conformations – of particular importance to polymer, peptide and protein structures.

Finally, this framework provides a new path for how ML can be used to address the limitations of experiment and theory by combining them together. The ML refinement process establishes a quantitative metric for assessing and optimizing DFT models by systematically exploring parameters such as basis sets, functionals, and core hole constraining approaches. In a complementary direction, the framework opens new avenues for ML applications. [57,62,63] Importantly, the clustering framework accepts TDMs from any computational methodology. The cluster-derived representations form a concise basis for transfer learning across related chemical families, while the clustered manifolds can serve as interpretable conditioning variables in generative models designed to produce molecules with specific spectral signatures. [64,65]. Furthermore, the energy-overlap distance metric employed here could be expanded in the future, for example, to incorporate TDM orientation or chemical bonding type. For more flexible molecules or less clearly differentiated atomic environments, such metrics could be combined with conformationally averaged calculations for improved fidelity. These synergistic directions could accelerate the discovery of materials with tailored optical properties and novel electronic states.

In conclusion, we have developed an ML algorithm to cluster DFT-calculated molecular transitions into an optical model refinable to polarized NEXAFS spectroscopy measurements. The result is a quantitatively accurate optical tensor describing polarized resonant X-ray interactions with that molecule. This model connects experimental resonance features directly to chemical moiety, MOs, and TDM orientations within the molecule and has been validated by reproducing NEXAFS of a different crystalline structure and orientation. Our approach achieves quantitative agreement between DFT and experimental data within 5% across the entire absorption edge, effectively bridging the gap between theory and experiment that will improve the accuracy of theoretical models. This method further enables label-free quantitative chemical and orientational analyses of individual moieties within molecular nanostructures using advanced resonant X-ray techniques, ultimately advancing diverse fields involving organic nanostructures.

**Acknowledgements** This work was supported by the Department of Energy Early Career Research Program under Grant DE-SC0017923. Funding support for OA was provided by Prince Satam bin Abdulaziz University, Saudi Arabia. Researcher HH was supported in part by an ALS Doctoral Fellowship in Residence**.** This research used resources SST-1 (7-ID-1) of the National Synchrotron Light Source II, a U.S. Department of Energy (DOE) Office of Science User Facility operated for the DOE Office of Science by Brookhaven National Laboratory under Contract No. DE-SC0012704. This research used resources of the Advanced Light Source, which is a DOE Office of Science User Facility under contract no. DE-AC02-05CH11231.

**Data Availability Statement** All data for this study will be made available upon reasonable request to the corresponding author. NEXAFS datasets are openly available on X-ray Atlas (CC BY 4.0). [51,59]

## Supporting Information

## Quantitative and bond-traceable resonant X-ray optical tensors of organic molecules

Victor Murcia[1], Obaid Alqahtani[1-3], Harlan Heilman[2], Brian A. Collins[1,2*]
*[1]Department of Materials Science and Engineering, Washington State University, Pullman, Washington 99164, USA*
*[2]Department of Physics and Astronomy, Washington State University, Pullman, Washington 99164, USA*
*[3]Department of Physics, Prince Satam bin Abdulaziz University, Alkharj 11942, KSA*

## Detailed Methods:

**Sample Growth:** Small molecule CuPc, ($C_{32}H_{16}CuN_8$ from Hui Chem Co Ltd.) with purity ≥99.5% was further purified in the physical vapor deposition (PVD) vacuum chamber (base pressure $< 1 \times 10^{-7}\ Torr$) by gradually heating, approaching CuPc sublimation temperature over several hours until no further outgassing of contaminants were detected by quartz crystal monitor (QCM). Thin films ($100\ nm$ measured by QCM) of CuPc were deposited at 0.25 Å/s on different substrates via PVD with the source held at a temperature of 375°C. The substrates were either conductive Si wafers (University Wafers) for GIXRD measurements and electron yield NEXAFS spectroscopy and $Si_3N_4$ windows for transmission NEXAFS spectroscopy. $Si_3N_4$ windows were composed of 100 nm thick $Si_3N_4$ membranes on a windowed (2x2 mm) Si frame (Noracada NX5200C). CuPc was either deposited directly on these substrates to create the herringbone crystal structure or onto a Copper(I) iodide (CuI, 3 nm) orientation patterning layer (deposited via PVD) to create the brickstone crystal structure. Both CuPc and CuI depositions were each accomplished in one single deposition on all samples for identical deposition conditions and thicknesses.

**NEXAFS Measurements:** The NEXAFS measurements were carried out at the SST-1 (7-ID-1) beamline [66] of the National Synchrotron Light Source (NSLS-II) in transmission mode on SiN window samples and total electron yield mode on Si wafer samples. Four different sample alignments with respect to the incident x-ray were used, namely 40, 55, 70 and 90 degrees. X-ray linear polarization was held constant in a direction that caused it to vary relative to the substrate normal as the sample was rotated. NEXAFS was collected over the energy range of 270-350eV via a photodiode and an Au mesh monitor upstream in the beamline. All four samples were probed via this methodology plus a version of the samples that only lacked the CuPc layer. Highly oriented pyrolytic graphite was also measured for energy calibration. All data was double normalized, processed in the QANT program, [67] and datasets used for clustering are available on X-ray atlas. [51,59,68] For transmission mode data, Beer's law was used to construct the spectra with the incident intensity based on transmission through the substrate-only version of each sample. This removed background absorbance from the $Si_3N_4$ membrane and the CuI layer. Comparison of the transmission data to the Si substrate data ensured identical NEXAFS signals on the Si substrate sample versions that were measured via GIXRD. Data uncertainties were estimated by averaging the AR-NEXAFS spectra across replicates and calculating the standard deviation at each energy. The standard error of the mean was then computed to reflect the uncertainty in the averaged spectrum. A single constant uncertainty value, obtained by averaging the standard error across all energies, was used in the chi-squared fitting procedure.

**GIXRD Measurements:** The GIXRD measurements were performed at beamline 7.3.3 [11] at the Advanced Light Source (ALS) in the Lawrence Berkeley National Lab (LBNL) set to an energy of 10keV and an incident angle of 0.2 degrees which is above the critical angle of both the films and substrates. The data was processed in Igor Pro by NIKA. [69]

**Molecular Geometry:** The molecular geometry for CuPc was extracted from a crystallographic information file (.cif) based on a 2003 XRD study. [49] DFT calculations were performed on a single isolated molecule and no additional structural or geometric calculations were performed on the nuclear ground state. The structure was translated such that the Cu atom coincided with the origin and the molecule was co-planar with the x-y coordinate plane with the x and y-axis aligned along the Cu-N bonds as shown in **Figure S1**. Chemically unique carbon atoms were chosen based on the molecules $D_{4h}$ symmetry and used as excitation centers in the DFT calculation. To account for the mixed ligand – covalent nature of the Cu-N bonds, 16 C atoms located in adjacent isoindole moieties were used as excitation centers (atoms C1-C16 in **Figure 2a**).

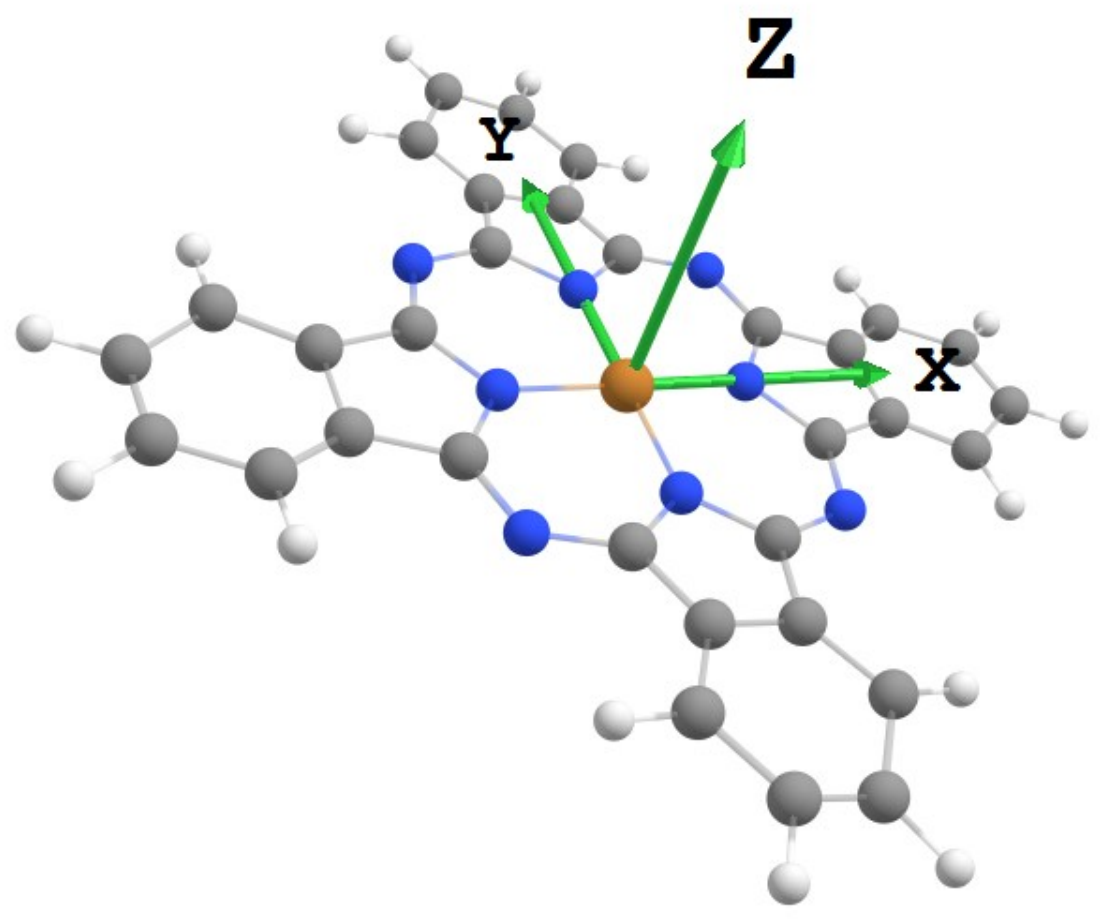

*Figure S1: Schematic of the CuPc molecule with the molecular reference frame axes superimposed.*

**DFT Calculations:** Orbitals, transition energies, and transition dipole moments were computed with the software package StoBe [70], combining Kohn-Sham total energy differences ($\Delta KS$) with a transition potential (TP) calculation. For each of the sixteen carbon excitation centers, three calculations were performed in the following order. First, a ground state SCF calculation fixed the total ground-state energy $E_i^g$ and identified the core 1s orbital of excitation center $i$. Second, an excited state calculation, placed an electron in the LUMO, while leaving a full core hole at the excitation center's 1s state, and it's total excited energy, $E_i^e$ defines the difference $E_i^e - E_i^g$ as the core-to-LUMO transition energy in the $\Delta KS$ sense. Third, the TP calculation was performed in which the same 1s orbital is held at half occupancy under the half-core-hole (HCH) approximation [71], consistent with prior transition-potential NEXAFS studies [31,54,72,73]. The TP calculation provides the full manifold of X-ray transitions with their oscillator strengths and transition dipole moments, which define the NEXAFS spectrum for each excitation center. It also yields, through Slater's transition state estimate, the ionization potential, the energy to remove the C 1e electron at that center to the continuum. [74]

The excitation-center carbon is described by the all-electron IGLO-III basis set [75], which resolves the C 1s orbital and localizes the core hole on that atom. Following our prior work, every other, non-excitation carbon is described by a modified valence basis (In StoBe O-CARBON(+4) (311/211/1)) together with a model core potential (MCP) that replaces its core s states, keeping the core hole unambiguous, which is essential for a well-defined NEXAFS calculation. Nitrogen and copper are described by all-electron triple-zeta valence (TZVP) basis sets and carry no model core potential, because their 1s levels lie far in binding energy from the carbon K-edge and offer no core-hole delocalization pathways at the C 1s excitation centers. Hydrogen is described by an all-election double zeta valence potential (DZVP) basis set. Auxiliary basis sets for all atoms use an A5 parameterization, with at least five s and p fitting functions per atom. The exchange-correlation functional is a generalized-gradient approximation (GGA), RPBE exchange [76] with PBE correlation [77]. These bases sets and functionals are chosen based on their alignment with other TP DFT studies of NEXAFS [54,72,73] shown to represent core and valence states in molecules of similar composition, although similar spectra can be generated with a different set of choices. Some optimization was performed on these choices [78] and the final level of theory nears the limit of what StoBe can provide.

The TP transition energies are accurate in relative space but not on an absolute scale, so the manifold of each excitation center $i$ is rigidly shifted onto a common absolute scale [79,80] by

$$E_i^c = E_i^e - E_i^g - E_i^l \tag{S1}$$

where $E_i^e$ and $E_i^g$ are the core-excited and ground state total energies, and $E_i^l$ is the lowest TP-calculated X-ray transition energy for center $i$, that is the core-to-LUMO transition computed within the HCH-TP scheme. $E_i^c$ is therefore the calibration offset between the TP and $\Delta KS$ estimates of the same core-to-LUMO transition. The same energy correction $E_i^c$ is applied uniformly to the entire TP manifold for each excitation center, including the ionization potential to place it on the same absolute scale. The transition dipole moments, molecular orbitals, and resonant energies reported in the main text follow from these ground, core-excited and transition-potential calculations for all sixteen carbon centers.

Unoccupied Molecular Orbitals were rendered in Chemcraft [81].

The clustering and fitting procedures were implemented using a custom codebase developed primarily in Igor Pro 8, with supplementary routines in Python for data handling and analysis. The source code can be found on our GitHub: https://github.com/WSU-Carbon-Lab/dft-

learn/releases/tag/v1.1 with more development available on the main branch to further automate the DFT calculation. [82]

**Calculating Bare Atom Absorption Step Edge from DFT**

The absorption step edge for CuPc was determined from the ionization energy calculated by DFT for each Carbon atom that served as an excitation center for the calculation. The ionization edge line shape for each excitation atom is assumed by the error-function functional form [12] with a decay fit to the bare atom mass absorbance of the molecule in the pre and post edge regions as calculated from the Henke database [83].

$$I_{DFT,Step} = H\left[\frac{1}{2}+\frac{1}{2}\mathrm{erf}\left(\frac{E-P}{\frac{\Gamma_G}{c^*}}\right)\right]e^{\beta\left(E-P-\frac{\Gamma_G}{c^*}\right)} \quad \text{(S2)}$$

$H$ is the height of the function after the step, $P$ is the position of the inflection point of the lineshape which will be defined by the IP, $\Gamma_G$ is the FWHM of the step defined by the broadening scheme used over the course of this work, $E$ is the photon energy, $c^*$is convolution factor for a Gaussian equal to $c\sqrt{\ln 2}$ where $c = 1.665$, and $\beta$ is an exponential decay factor. The total DFT step edge is then the coherent sum of each of the step edges constructed from each excitation center. Finally, the total DFT step edge is scaled to the bare atom mass absorption constructed from the Henke database using Min-Max scaling at pre-edge (below 283 eV) and post-edge (above 320 eV) energies. **Figure S2** shows the DFT-calculated step edge superimposed on the Henke Bare Atom mass absorbance data.

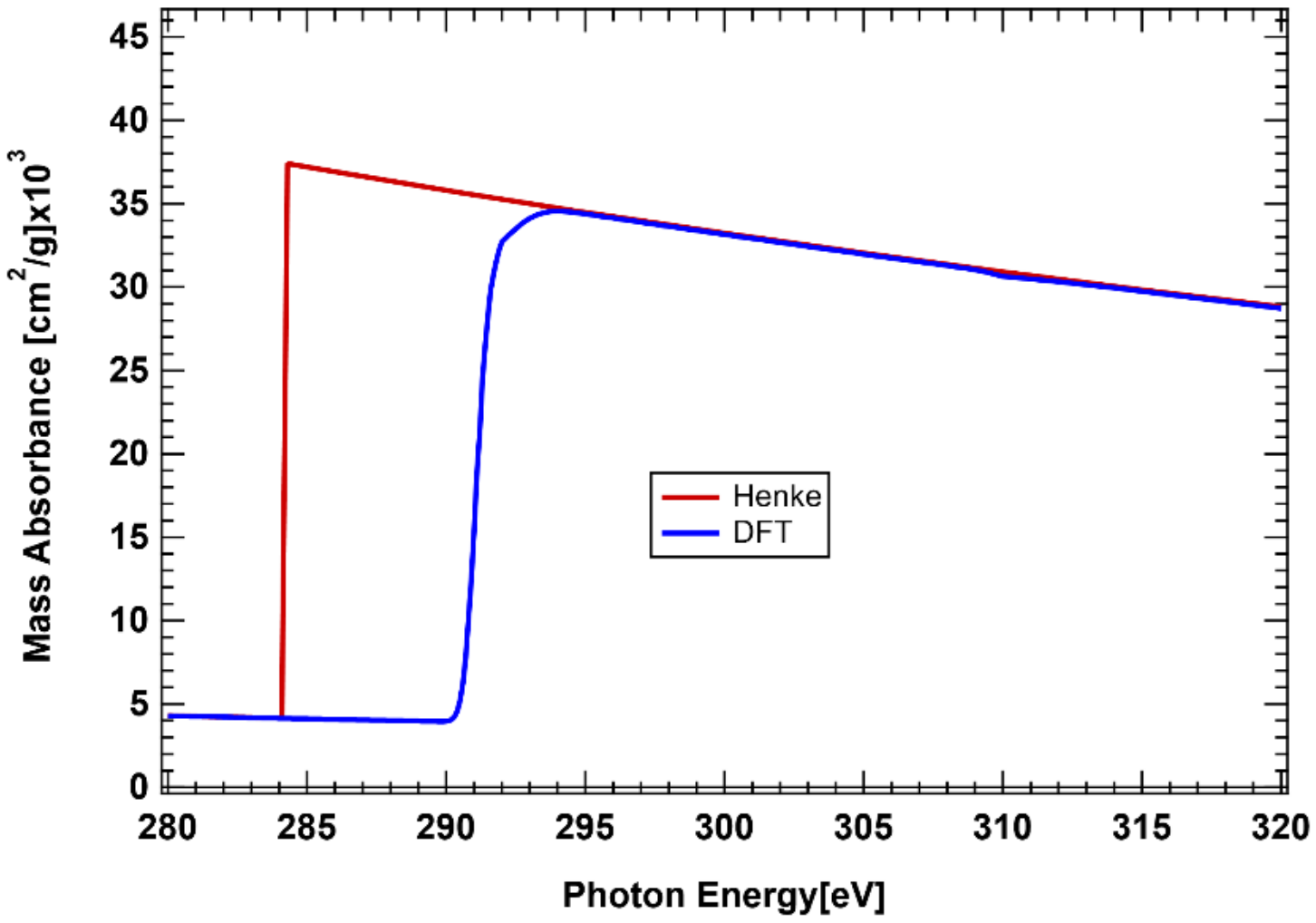

*Figure S2: Comparison between the Henke bare atom mass absorption spectrum for CuPc vs the DFT-modified version based on the IP of each excitation center in the molecule.*

**Spectral Broadening**: The energy dependence of the mass absorbance spectra is determined by applying a linearly interpolated energy broadening of 0.5 to 12 eV across the carbon K-edge (285 to 320 eV) using a gaussian line shape as described previously in section 7.2.2 of [12] and further detailed in the StoBe manual (Section 2.7.6) [70]. More specifically, below the IP of each excitation center, the broadening is a constant 0.5eV and above 320 eV the broadening is a constant 12 eV. The spectral range was chosen to encompass both $\pi^*$ and $\sigma^*$ transitions present in common experimental data.

**Calculating Mass Absorbance and Optical Tensor from DFT:** DFT calculations are done in the molecular frame of reference for a CuPc molecule approximately co-planar with the $x, y$-plane. The dielectric tensor for a single molecule is calculated from the TDM's that determine the imaginary part of the tensor which is sufficient to describe NEXAFS spectroscopy.

To model a uniaxial CuPc film, the single-molecule dielectric tensor is first rotated by a polar angle $\alpha$ corresponding to the molecular tilt angle between the film surface and the molecular plane. The uniaxial tensor is then constructed from the ensemble average over azimuthal – in plane rotations. The resulting film tensor components $\chi^f$ are expressed as

$$\chi^f_{xx}(\alpha) = \chi^f_{yy} = \frac{1}{2}(\chi_{xx}(1 + \cos^2 \alpha) + \chi_{zz} \sin^2 \alpha) \tag{S3}$$

$$\chi^f_{zz}(\alpha) = \chi_{xx} \sin^2 \alpha + \chi_{zz} \cos^2 \alpha \tag{S4}$$

where $\chi^f_{ii}(\alpha)$ is the index of refraction along the $ii^{th}$ material axis of the film, and $\alpha$ can be interpreted as the molecular tilt angle or cone angle.

To further account for molecular orientation, the tensor is rotated about the molecular x-axis by the tilt angle $\alpha$ (which has been empirically determined) along the molecular x-axis producing a tilted transition tensor:

$$\chi_{Tilt}(E_i, \alpha) = R_\alpha(\alpha)\chi_N(E_i)R^T_\alpha(\alpha) \tag{S5}$$

This tensor is subsequently rotated in a fourfold manner about the surface normal (z-axis) to approximate the radial symmetry of the film. The rotated tensors are then summed to obtain the effective film tensor associated with each transition (or cluster of transitions):

$$\chi_{Film}(E_i, \gamma_j) = \sum_{j=1}^{4} R_z(\gamma_j)\chi_{Tilt}(E_i, \alpha)R^T_z(\gamma_j) \quad \gamma_j = 0°, 90°, 180°, 270° \tag{S6}$$

The NEXAFS signal is calculated by projecting the film tensor onto the incident electric field vector, which is defined by the experimental geometry through the sample angle $\theta$. For the measurements considered here, $\theta$ takes values of $40°, 55°, 70°$ or $90°$. Since we are using linearly polarized light, $\phi = 0°$ for all sample alignments.

$$E = \begin{pmatrix} \sin\theta \cos\phi \\ \sin\theta \sin\phi \\ \cos\theta \end{pmatrix} \tag{S7}$$

The result of impinging the tensor via an inner product with the electric field, results in a contraction of the tensor down to a scalar value that represents the mass absorbance:

$$\mu_{MA} = E\chi_{Film}E^T \qquad \text{(S8)}$$

The clustered NEXAFS spectrum is constructed by summing the contributions from all clusters, each characterized by an energy, amplitude, width, and orientation derived from the underlying DFT transitions. A global scaling factor is applied to match the overall intensity of the experimental data. The molecular tilt angle α and the cluster-specific parameters collectively determine the angular and spectral dependence of the modeled response:

$$I_{NEXAFS,Film} = i_0 \sum_{j=1}^{4} \sum_{i}^{N} R_\alpha(\alpha) R_Z(\gamma_j) \chi_{N.i} R_Z^T(\gamma_j) R_\alpha^T(\alpha) \Gamma(E_i, w_i, f_i, \theta_{Mod,i}) \qquad \text{(S9)}$$

where $i_0$ is global scaling/proportionality factor between the clusters and the experiment, $\alpha$ is the molecular tilt angle, $\gamma_j$ is one of the 4 rotation angles used to simulate the film tensor (i.e., $0°, 90°, 180°$, and $360°$), $E_i$ is the position of cluster $i$, $w_i$ is the width of cluster $i$, $f_i$ is the amplitude of cluster $i$ and $\theta_i$ is the orientation of cluster $i$ determined from ratio of the diagonal tensor elements $xx$ and $zz$.

**Clustering Algorithm for TP-DFT NEXAFS Transitions:**

First low oscillator strength transitions are filtered out via an oscillator strength threshold (OST) in order to preserve the strongest transitions. **Figure S3** shows this process where the grey sticks on the graph to the right shows the transitions filtered out lying below an OST of 2%.

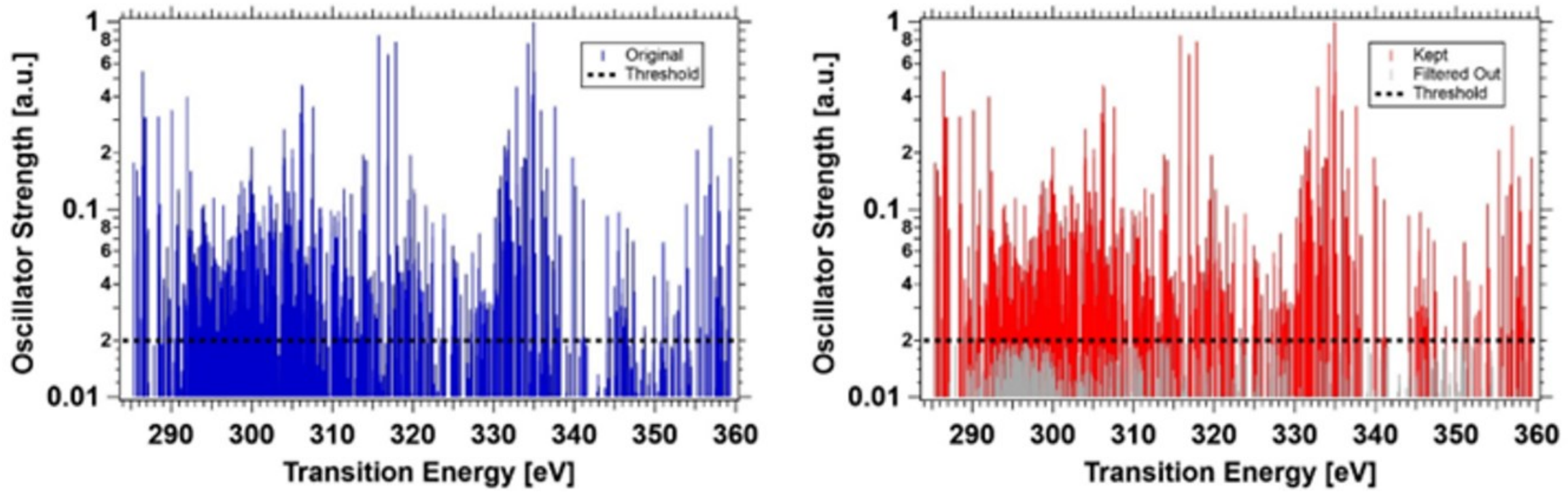

*Figure S3:. Example of effect of the oscillator strength threshold (OST=2%) where the original DFT-calculated transitions (left) are filtered such that those with an oscillator strength <2% are eliminated (grey sticks on the right graph).*

Each transition is modeled as a Gaussian function where the position is defined according to the transition energy, the amplitude is defined by the oscillator strength and the width by a piecewise linear energy-dependent broadening scheme described above [12]. With this Gaussian representation, we define the amount of overlap between each transition pair as:

$$\%OVP = \frac{F_i(c) + F_j(c)}{\min(A_i, A_j)} \times 100 = \frac{A_{OVP}}{\min(A_i, A_j)} \times 100 \qquad \text{(S10)}$$

where $F_i(c)$ and $F_j(c)$ are the cumulative distribution functions (CDFs) for the ith transition and the jth transition evaluated at the intersection point "$c$" between the peaks, $A_{OVP}$ represents the combined overlap area between the two Gaussian peaks, evaluated at their intersection point, and $\min(A_i, A_j)$ is the area of the smallest transition peak in the pairing.

The CDFs for the Gaussian's are computed as shown below:

$$F_i(x) = \int f_i(x)dx = \frac{1}{2}a_i\sigma_i\sqrt{2\pi}\left(1 + erf\left(\frac{(\mp\mu_i \pm x)}{\sigma_i\sqrt{2}}\right)\right) \quad \text{(S11)}$$

where $f_i(x)$ represents the Gaussian function of the ith transition, $a_i$ represents the oscillator strength, $\sigma_i$ represents the width in terms of the standard deviation, and $\mu_i$ is the transition energy.

To determine the intersection point $c$, we set the Gaussian functions for each transition equal to one another and solve for "$x$". When $\mu_i < \mu_j$ (energy of the gaussians are in ascending order $c$ is determined by the following equation:

$$c = \frac{\mu_j\sigma_i^2 - \sigma_j^2\left(\mu_i\sigma_j + \sigma_i\sqrt{\left(\mu_i - \mu_j\right)^2 + 2\left(\sigma_i^2 - \sigma_j^2\right)\log\frac{\sigma_i}{\sigma_j}}\right)}{\sigma_i^2 + \sigma_j^2} \quad \text{(S12)}$$

This enables us to determine the $\%_{OVP}$ for every transition pair in order to produce the peak overlap matrix. The overlap matrix can then be used to cluster transitions depending on whether their $\%_{OVP}$ is above or below the peak overlap threshold (OVPT). If a transition pair has a $\%_{OVP}$ above the OVPT then those transitions are considered to have a sufficiently high energetic overlap and can thus be placed in a cluster together. If a transition pair has a $\%_{OVP}$ below the OVP% then those transitions are considered to not have a sufficiently high energetic overlap and are considered to be different and don't belong in a cluster together.

The clustering process, illustrated in **Figure S4** begins by evaluating the overlap matrix, which quantifies the percentage overlap ($\%_{OVP}$). Starting from the first row and first column, the algorithm checks whether the $\%_{OVP}$ exceeds the OVPT (**Figure S4b)**. If so, the corresponding transition is added to the current cluster. The algorithm then proceeds along the same row, including transitions with $\%_{OVP}$ values above the threshold (**Figure S4c)**. When a $\%_{OVP}$ falls below the OVPT for the first time, the algorithm skips to the next unprocessed transition and resumes the check. If a second below-threshold $\%_{OVP}$ is encountered, the algorithm identifies the matrix element with the highest oscillator strength among the current transitions in the cluster and uses it to update the row and column indices (**Figure S4d)**. The algorithm continues along this row until a transition with $\%_{OVP}$ below the OVPT is reached (**Figure S4e)**. All transitions with $\%_{OVP}$ above the threshold processed so far represent the current cluster (**Figure S4f)**. The algorithm resumes from the earliest unclustered transition. This procedure continues until all transition pairs have been evaluated and assigned to clusters.

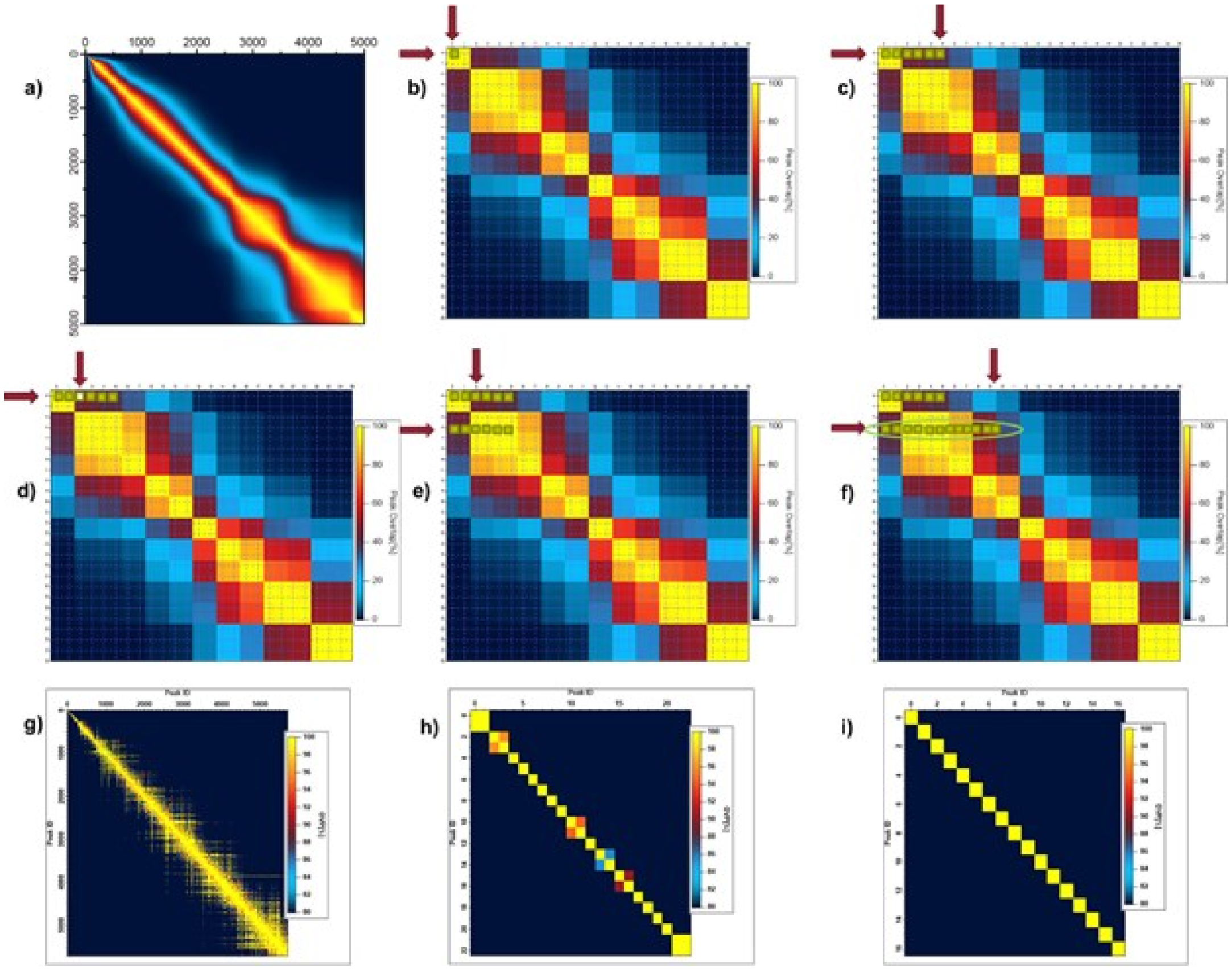

*Figure S4. Visualization of the clustering algorithm. Plot a) shows an overlap matrix. The axes of the overlap matrix correspond to transition indices, with each element representing the percentage overlap (%OVP) between a pair of transitions. Plots b-f) is a zoom of the overlap matrix and shows how a cluster is generated (see above text). Plots g-i) show the result of the clustering process at different iterations*

Now that the transitions have been grouped into a set of clusters based on their energetic overlap, we combine the transitions comprising each cluster into a Gaussian representation that has its own position, width and amplitude that is still capable of reproducing the DFT NEXAFS. Each transition that makes up the cluster is described by a Gaussian peak

$$g_{pk} = \frac{A}{\sigma\sqrt{2\pi}} e^{-\frac{1}{2}\left(\frac{x-\mu}{\sigma}\right)^2} \quad \text{(S13)}$$

where $A$ is the peak amplitude defined by the oscillator strength $(f_i)$ calculated from DFT, $\sigma$ is the peak width defined by the broadening scheme described earlier, and $\mu$ is the peak position defined by the transition energy calculated from DFT. The exponential prefactor $\frac{A}{\sigma\sqrt{2\pi}}$ defines the height of the Gaussian. The transitions in the cluster can be combined by adding all the Gaussians together to make a "summed peak", $S_{pk}$.

$$S_{pk} = \sum_{i}^{N} g_{pk,i} = \sum_{i}^{N} \frac{f_i}{\sigma_i\sqrt{2\pi}} e^{-\frac{1}{2}\left(\frac{x-\mu_i}{\sigma_i}\right)^2} \quad \text{(S14)}$$

The width of the merged peak ($\sigma_M$) will be defined by the full width half maximum of the summed peak. The amplitude of the merged peak ($f_M$) will be defined in terms of the height ($H$) of the summed peak as $H\sigma_M\sqrt{2\pi}$ (**Figure S5**).

$$M_{pk} = \frac{f_M}{\sigma_i\sqrt{2\pi}} e^{-\frac{1}{2}\left(\frac{x-\mu_M}{\sigma_M}\right)^2} \tag{S15}$$

Lastly, the TDM associated with each cluster is obtained by summing the TDM components of the member transitions. Since the magnitude of each TDM is proportional to the corresponding of the oscillator strength, this construction inherently incorporates intensity weighting, such that stronger transitions contribute more strongly to the resulting cluster dipole. The cluster orientation angle $\theta$ is then recomputed from the resulting cluster TDM relative to the molecular z-axis. The transition tensor for the cluster is then calculated using the components $s_x, s_y, s_z$ of the cluster TDM via the transform:

$$\tau_{Mol} = R\tau_{TDM}R^T = \begin{pmatrix} s_x^2 & s_x s_y & s_x s_z \\ s_y s_x & s_y^2 & s_y s_z \\ s_z s_x & s_z s_y & s_z^2 \end{pmatrix} \tag{S16}$$

where $R$ is the 3D rotation matrix.

The merged representation is generated for each cluster, and a peak overlap matrix is produced for them. If the peak overlap matrix contains any nondiagonal elements that exceed the OVPT, the clustering routine is done on the merged peaks. The process is iteratively repeated until a peak overlap matrix that has no elements exceeding the OVPT is obtained. **Figure S4(g-i)** shows the result of applying this clustering process over three iterations.

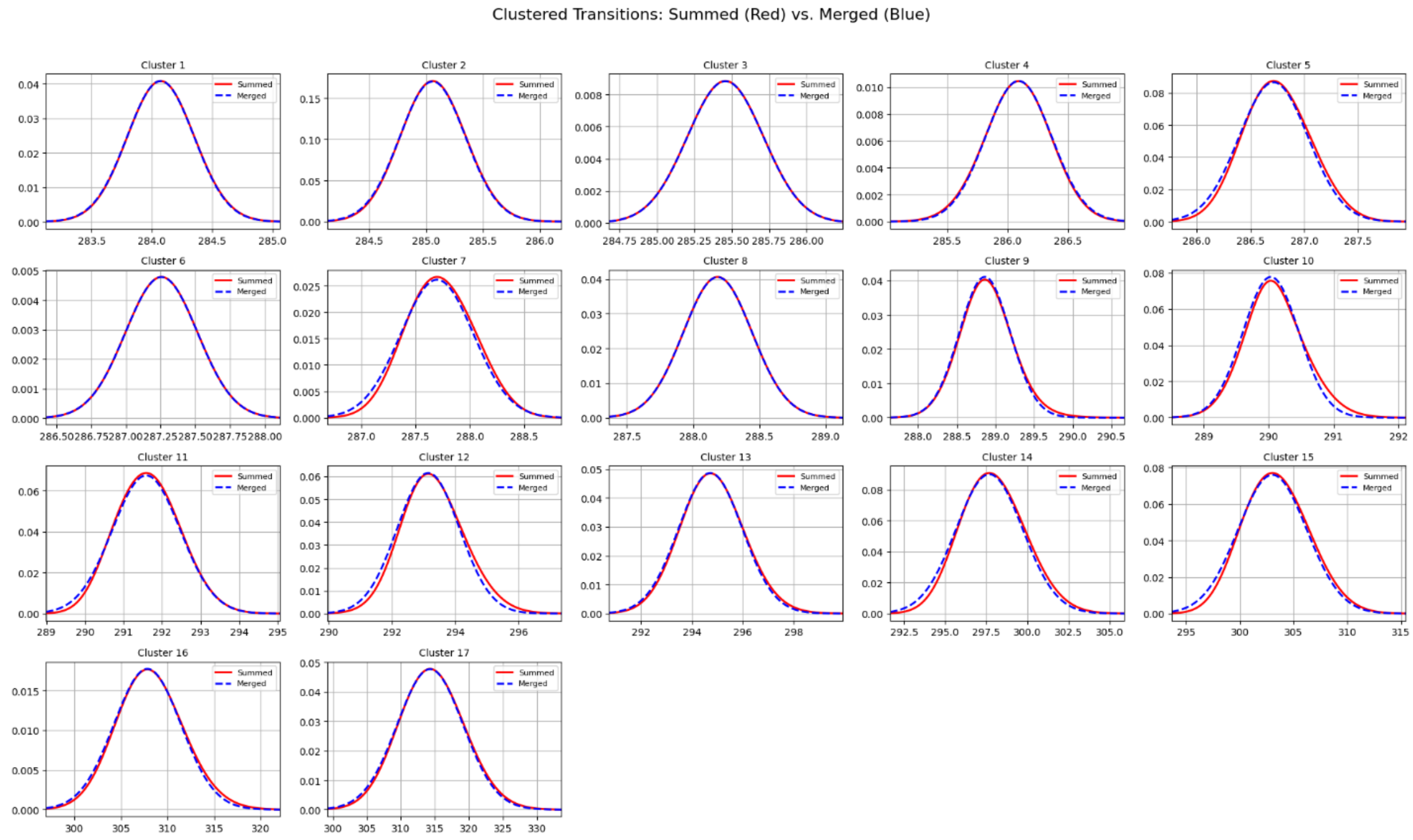

*Figure S5. Comparison between summed and merged peaks for the final clustered DFT model used in this work.*

## Simultaneous Fitting of Angle-Resolved NEXAFS

The following steps detail the process to refine the DFT cluster model to the angle-resolved NEXAFS spectroscopy measurements. This process involved several steps with increasing degrees of freedom applied to the clustered DFT model. These approaches are applied sequentially, with each stage building upon the previous fit.

The steps are as follows with example outputs from step 3 shown in **Figure S6** below.

1. Scale the DFT clustered model to the bare-atom absorption scaled NEXAFS via $i_0$
2. Determine the molecular tilt angle $\alpha$ from angle-resolved NEXAFS
3. Refine the DFT cluster model using fitting approaches with increasing numbers of free parameters:
    a. Peak amplitudes are treated as free fitting parameters; all other parameters are fixed.
    b. Peak amplitudes and transition dipole moment (TDM) orientations are treated as free fitting parameters; all other parameters are fixed.
    c. Peak amplitudes, positions, and TDM orientations are treated as free fitting parameters; all other parameters are fixed.
    d. Peak amplitudes, positions, widths, and TDM orientations are treated as free fitting parameters, while $\alpha$is fixed.
    e. All model parameters are treated as free fitting parameters.

We begin by determining a scaling factor between the DFT cluster model and the bare atom scaled NEXAFS. This is achieved through the parameter $i_0$, which acts as a multiplicative factor. To determine $i_0$, a fit to the first $\pi$-manifold peaks is performed by allowing $i_0$ to vary while the reduced chi squared ($\chi^2_{red}$) statistic used as a metric of goodness of fit in these routines is minimized.

The process for determining $\alpha$ involves first selecting an energy range that will serve as the basis for the orientational analysis. Within this range, the NEXAFS intensity for every energy point will be extracted for every sample alignment. Then, the intensities will be plotted as a function of sample alignment for a given energy which allows the tilt angle to be extracted via:

$$I_v^{\parallel} = \frac{1}{3}\left[1 + \frac{1}{2}(3\cos^2\theta - 1)(3\cos^2\alpha - 1)\right] \quad \text{(S17)}$$

This value of $\alpha$ serves as a fit parameter that will be refined in the final stage of the fitting routine.

Finally, every cluster in the optical model will be represented by a peak function (a Gaussian in this case). Each of these Gaussians will have four parameters tied to it. The parameters are the peak position, the peak width, the peak amplitude, and the peak orientation. In addition to this, the global molecular orientation defined by $\alpha$ will also be part of the final stage the fit. Therefore, the model will have a total of $4N + 1$ parameters, where $N$ is the number of clusters in the DFT-BB model. The parameters are constrained as follows:

- The amplitudes are constrained to be positive.
- The $\theta_{Mod}$ is constrained to be within 0° and 90°.
- The positions are constrained to stay within $\pm 2 * FWHM$ in an attempt to prevent the clusters from overlapping too much and prevent the cluster order from being disrupted.
- The widths are constrained to not change by more than twice their initial value.
- The tilt angle is constrained to not change by more than 5° from the initial value

Among these approaches, the selected fitting method provides the best balance between quantitative agreement with experiment and minimal deviation from the original DFT-derived parameters, as discussed in the main text.

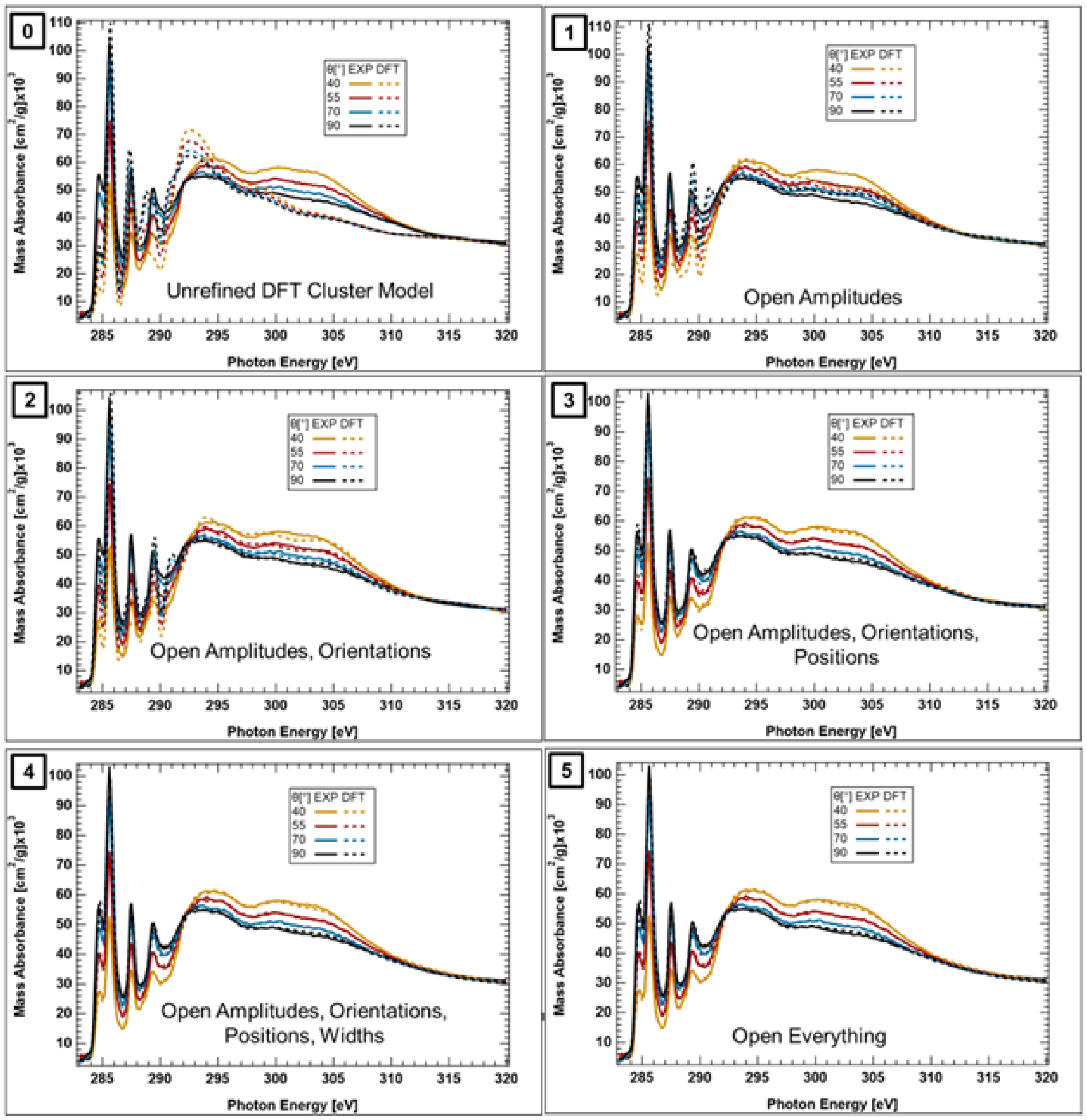

*Figure S6. Different fitting stages for refinement of DFT cluster model to AR-NEXAFS. The model is shown as dashed traces while the experiment is shown as solid traces. 0) Unrefined cluster model 1) Free amplitudes 2) Free amplitudes, orientations 3) Free amplitudes, orientations, positions 4) Free amplitudes, orientations, positions, orientations, widths 5) Free amplitudes, orientations, positions, widths and molecular tilt angle.*

<u>**Supplementary Graphs:**</u>

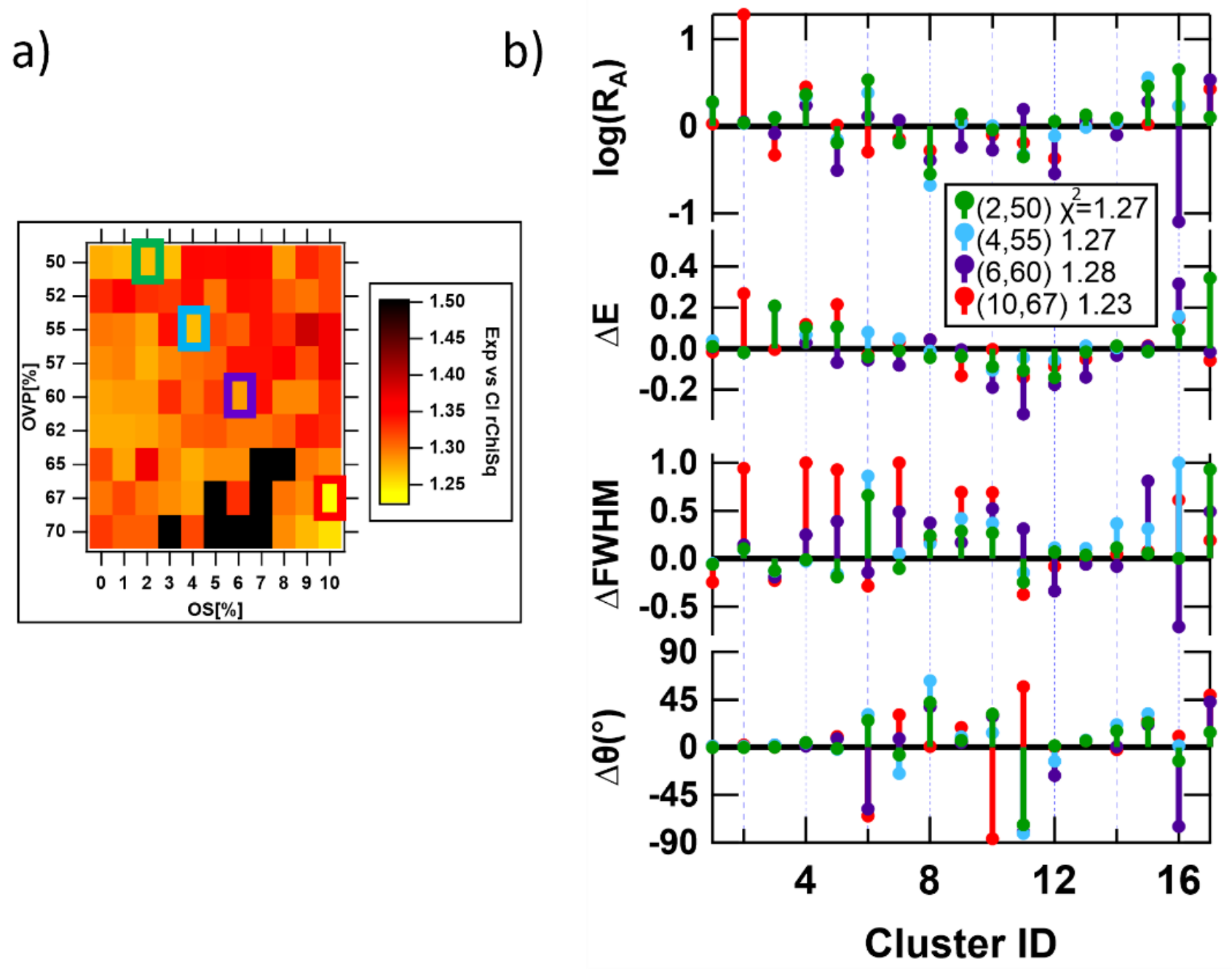

*Figure S7. Identifying the best clustering model based on low $\chi^2$ and minimal changes to parameters upon refinement to experiment. a) $\chi^2$ as a function of OST and OPVT with fits from four candidate clustering models framed. (All models were considered.) b) Change in model parameters after refinement for the four candidate clustering models color coded by the frames in (a). The legend shows the (OST[%],OVPT[%]) and the $\chi^2$ for each fit. The chosen set (OST=2% & OVPT=50%) minimized the magnitude of cluster parameter changes necessary to fit the experimental data.*

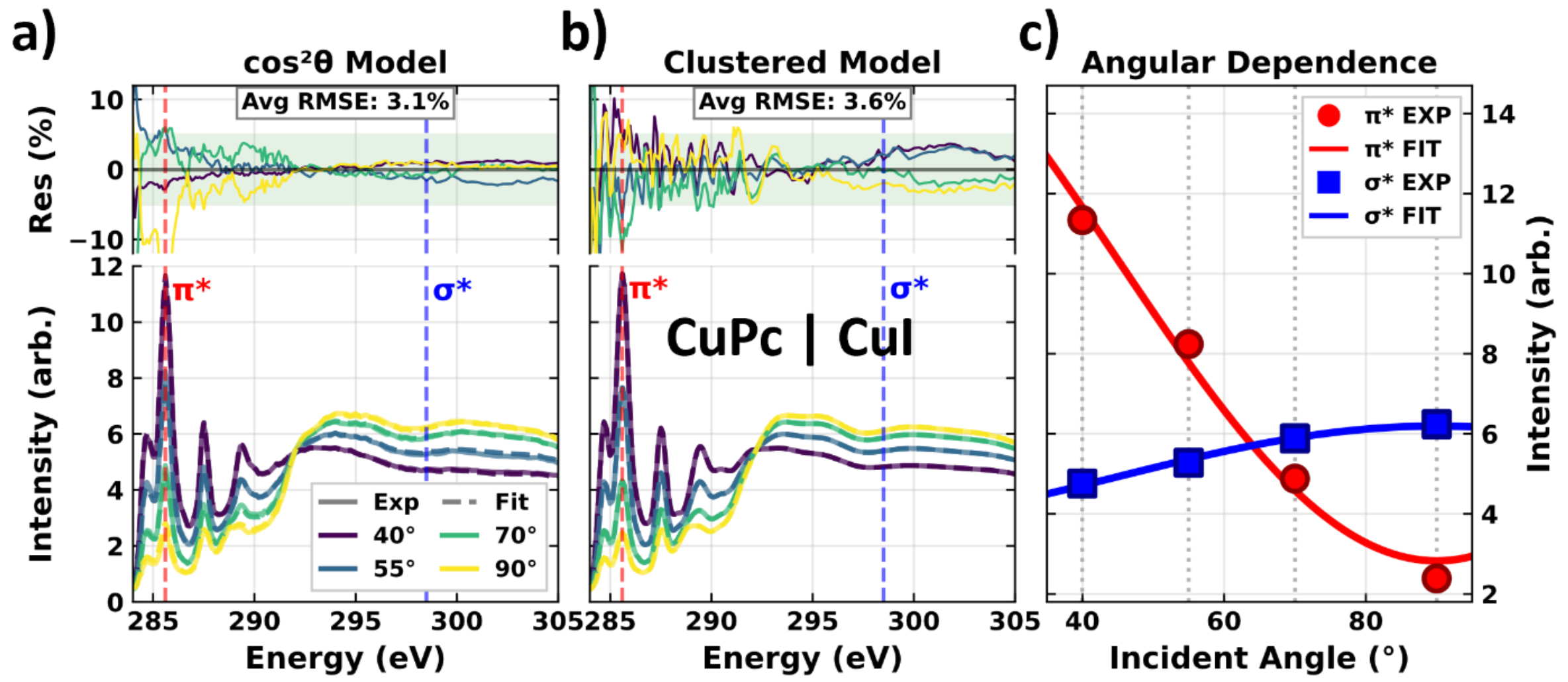

*Figure S8. Evaluating the goodness of fit of the optical model to the brickstone CuPc sample against that of the empirical $\cos^2\theta$ model. (a) Result of fitting angle resolved NEXAFS spectra at every measured energy separately to a $\cos^2\theta$ function. Top are residuals of the fit where the highlighted green portion correspond to residuals within 5%. (b) The same for our DFT cluster-refined model. Note the RMSE of these two are nearly identical showing that the DFT model is as quantitatively accurate as the empirical model. (c) example fits of the $\cos^2\theta$ empirical model to experimental data at 285.6 eV (blue) and 298.5 eV (red) corresponding to transitions with pi and sigma characteristics as determined from clusters 2 and 14.*

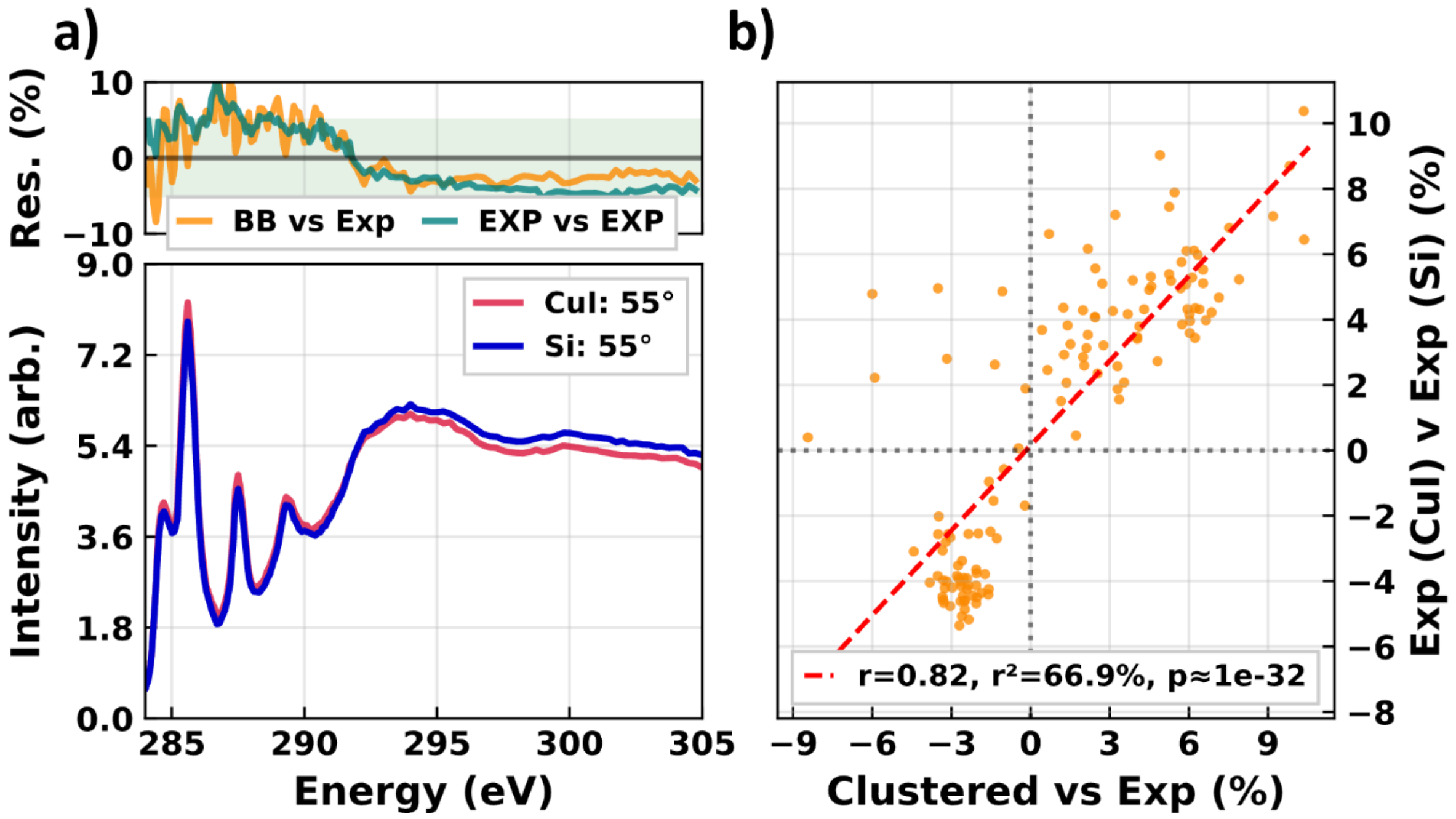

*Figure S9.(a) Comparison between the magic angle NEXAFS spectra collected for the two samples. Top graph is analysis of the percent residuals calculated between the two experimental spectra (green) as compared to the residuals calculated between the DFT model for CuPc on Si. (b) Correlation analysis between the model and experimental residuals showing that the errors (residuals) in the DFT model reproducing the herringbone NEXAFS spectra in Figure 5 of the main text. A Pearson correlation analysis reveals a strong, positive, and highly statistically significant linear relationship between the two sets of residuals. With a Pearson's coefficient ($r$) of .82 and a coefficient of determination ($r^2$) of 66.9%, over two-thirds of the variation in the DFT models error can be predicted by the variance between the two experimental samples. The extremely low p-value ($p$ ≈1e-32) confirms this relation is likely not due to random chance and supports the conclusion that a majority of the DFT model's error originates from the differences between experimental samples.*